# Quick Detection of Contaminants Leaching from Polypropylene Centrifuge Tube with Surface Enhanced Raman Spectroscopy and Ultra Violet Absorption Spectroscopy

Zhida Xu[1*], Manas R. Gartia[2], Charles J Choi[1], Jing Jiang[1], Yi Chen[1],Brian T Cunningham[1,3], Gang Logan Liu[1,3]

[1]Department of Electrical and Computer Engineering, University of Illinois at Urbana-Champaign, Urbana, 61801, USA

[2] Department of Nuclear, Plasma, and Radiological Engineering, University of Illinois at Urbana-Champaign, Urbana, 61801, USA

[3]Department of Bioengineering, University of Illinois at Urbana-Champaign, Urbana, 61801, USA

*zhidaxu1@illinois.edu

## Abstract

Anomalous surface enhanced Raman scattering (SERS) peaks are identified for liquid sample stored in polypropylene centrifuge tubes (PP tube) for months. We observed the unexpected Raman peaks during experiments for Thiamine Hydrochloride aqueous solution stored in PP tube for two months. In order to identify the contaminants we have performed SERS experiments for de-ionized water (DI water) stored in polypropylene centrifuge tube for two months and compared them with fresh DI water sample. We have also carried out Ultra Violet (UV) absorption spectra for both fresh and contaminated water. We believe that the water is contaminated because of chemicals leaching from the PP tube. From the GC-MS data the main contaminant was found to be Phthalic acid and its derivatives. Further SERS and UV absorption experiment for Phthalic acid correlates well with the anomalous peaks identified earlier. We qualitatively confirmed the identification and quantitatively estimated the concentration of suspect contaminants as between 1uM and 10uM  with both SERS and UV absorption spectroscopy. With UV absorption spectroscopy, we precisely estimate the concentration as 2.1uM.  We have shown that sample in PP tube can be contaminated due to leaching chemicals upon long term storage and suggested SERS and UV-absorption spectroscopy as two quick and simple techniques to detect the contamination.

Keywords:  Leaching,  Polypropylene  centrifuge  tube,  SERS,  UV-absorption spectroscopy, Mass spectroscopy

1. Introduction

**It has been known for long that some plasticware container can contaminate its contents and may be harmful to health. For example, bottle waters were found to be contaminated by antimony leaching from polyethylene terephthalate (PET).**[1]

**{{14 Shotyk,W. 2007}} The plasticizer bisphenol A (BPA) which is used in the production of polycarbonate is widely found in drinking water and is reported to stimulate cell apoptosis.**[2]

**{{15 Vandenberg,Laura N. 2007}} A sulfoxide oxidative product along with a sulfone oxidative of didodecyl 3,3′-thiodipropionate (DDTDP) that is used to prevent oxidative degradation of synthetic polymers leaching  from the**

**polypropylene tubes had been reported and the chemicals were identified with mass spectroscopy.** [3]

**{{12 Xia,Yuan-Qing 2005}} Plasticizers outgasing from o-rings can lead to undesired ion-molecule chemistry in an electrospray quadruple ion trap mass spectrometer.** [4]

**{{19 Verge,Kent M. 2002}} Recently it is reported that compounds such as di(2-hydroxyethyl)methyldodecylammonium (DiHEMDA) and 9-octadecenamide (oleamide) leaching from polypropylene tubes hamper the measurements in DNA and proteins assays.** [5]

{{13 McDonald,G.Reid 2008}}

**Even though in most cases the leaching compounds from plasticware are in negligible amounts, that are unlikely to be toxic to human body, but it may hamper experiment results, especially for those highly sensitive biochemical experiments. Due to the wider spread use of plastic containers both in daily life and in the laboratory, it is imperative to detect the contaminants leaching from the plasticware in a simple and fast way. In most of the previous works, mass spectroscopy was used to identify the leaching compounds. Reliable as it is, mass spectroscopy has some limitations. Usually the sample preparation to data acquisition processes in those instrument are time consuming and may take hours. Also, to get sufficient amount of leaching compounds, considerably large amount of sample solution needs to be vaporized. The mass spectrometer instrument tends to be complicated, huge and has special requirements for the operation environment and so it is difficult to be portable. To identify the structure of the compound, the molecules need to be break into fragments with certain techniques such as high energy electron bombardment.** [6]

{{21 Fenn,JB 1989}} Therefore the sample is not reusable.

**Surface enhanced Raman spectroscopy (SERS) can overcome these limitations. As a label free and non-invasive technique, Raman spectroscopy can tell information about vibrational mode and symmetry of a molecule and thus can identify the chemical species.** [7]

**{{22 K. D. Jernshoj, S. Hassing 2010}} Compared with normal Raman spectroscopy, SERS can enhance the Raman signal by several to tens order of magnitudes, which allows this spectroscopy technique to be sensitive enough to detect single molecules.** [8]

**{{23 Nie,Shuming 1997}} Ultraviolet absorption spectrometry is another simple and quick quantitative technique for molecular detection and identification but it is not as sensitive as SERS and may damage the structures of organic compounds due to its high photon energy.** [9]

{{24 Wang,Yunming 2010}}

Observations of anomalous Raman peaks in the SERS experiment of Thiamine Hydrochloride solution led us to examine the potential source of interferences from polypropylene centrifuge tubes (PP tube) as the solution was stored for about two months before the experiment. We observed those anomalous peaks at fixed wavenumbers and with constant intensity irrespective of the concentration of Thiamine Hydrochloride solution. For comparison we also took SERS spectra of freshly prepared Thiamine Hydrochloride solution with the same setup and configuration but did not observe those peaks. In order to confirm the leaching of PP tube, we perform SERS experiment for fresh deionized (DI) water and DI water stored in a similar PP tube for about two months (we call it old DI water). Those same anomalous Raman peaks showed up in old DI water while not in fresh DI water. Thus we assert that the water can be contaminated by PP tube and those anomalous Raman peaks can be attributed to the chemicals leaching from the PP tube. In addition, we took the ultra-violet (UV) absorption spectra of old DI water and fresh DI water. We observed two distinct absorption peaks in the spectrum for old DI water while not in fresh DI water. Further, in order to identify the source of interference leached from PP tubes, the old DI water samples were analyzed by GC-MS. The mass spectrometry data revealed that chemicals leached from PP tubes have low molecular weights (m/z < 500 Da; here m/z is the mass to charge ratio) and the major contaminants may be Phthalic acid. Finally, we have performed SERS and UV absorption experiments for Phthalic acid solutions. The characteristic peaks from Phthalic acid correlates well with the earlier observed anomalous peaks from old DI water. The details of the experiments are described in following sections.

2. Experimental configuration

2.1 Nanodome SERS substrate

**Figure 1 (a) shows the photograph of polymer based SERS substrate made using nano replica molding process. The details of the fabrication process is described elsewhere** [10]

**{{81 Charles J Choi,et al 2010}}. In brief, the SERS substrate used here is a two-dimensional periodic array of closely spaced plastic cylinders (nanodomes) coated by a thin layer of silver on the top. Figure 1 (b) shows the scanning electron micrograph of fabricated substrate after silver coating. The reproducibility and uniformity of this SERS substrate have already been demonstrated** [10]

**{{81 Charles J Choi,et al 2010}}, which allows us to get repeatable, reliable data for quantitative analysis. The enhancement factor for the substrate was calculated using our earlier developed method** [11]

{{105 Gartia,Manas R. 2010}} with Rhodamine-6G as SERS probe and found to be $3.16 \times 10^6$.

2.2 Spectroscopy setup for data acquisition

The schematic of our SERS spectroscopy system is shown in Fig. S1. For excitation, a semiconductor laser beam with the wavelength of 785nm and the power of 30mW is focused on to the SERS substrate by a 10X objective lens after reflected by a dichroic mirror. The diameter of the laser spot on the substrate is about 20um as measured by the camera.

For UV-absorption spectroscopy, we use Evolution 60 UV-Visible Spectrophotometer (Thermo Fisher Scientific). For mass spectroscopy, we use Agilent 6890N GC/5973 MS (GC/MS) gas chromatography mass spectrometer. It is equipped with electron impact ionization (EI) and chemical ionization units; it has flame ionization detector, thermal conductivity detector and mass selective detector.

2.3 Sample preparation

Thiamine Hydrochloride is dissolved in de-ionized (DI) water to make solutions of different concentrations. Then we contain the Thiamine Hydrochloride solutions and DI water sample in PP tubes and keep them in storage at room temperature for two months. For new samples, the Thiamine Hydrochloride with the same concentrations and fresh DI water are prepared right before the measurements. Glass containers are used for the new samples.

**For SERS spectral measurements, we drop the sample with the volume of 3 uL onto the SERS substrate. Due to the hydrophobicity of the substrate surface which results from the nanodome arrary structure,** [12]

{{27 Zhao,Wenjie 2010}} the drops stand on the substrate with the contact angle more than 90°. Then we wait a few minutes until it dries. For Thiamine Hydrochloride solutions of high concentration, a stain due to crystallization with the diameter of about 200um is left after drying. For low concentration Thiamine Hydrochloride solutions and DI water, nothing is visible to naked eyes after drying.

3. Experiment results and discussion

3.1 Interference of SERS spectra for old Thiamine hydrochloride solution

**As stated in the experimental section, the sample is prepared by dropping 3uL of liquid on the SERS substrate. All the spectrums presented here are captured with the integration time of 10 seconds. To facilitate the visualization and analysis of the Raman peaks, the autofluorescence background is removed with a modified multi-polynomial fitting algorithm.** [13]

{{16 Zhao,J. 2007}}

**Fig. 2(A) shows the SERS spectra for both old and new Thiamine hydrochloride (TH) solution with the concentrations of 1uM, 10uM and 100uM. The peak at 764 cm⁻¹ showing up on all curves must come from Thiamine hydrochloride since it varies with the TH concentration. We also confirm this peak with regular Raman spectra of 10mM TH solution (Fig. S2). The strong peak at 764 cm⁻¹ is generally attributed to the Pyrimidine ring breathing mode of TH.** [14]

{{552 Leopold,Nicolae 2005}} A closer comparison of 10 uM TH solution in old water revealed two anomalous peaks at 1008 and 1047 cm⁻¹ which did not show up for TH solution prepared with fresh water [Fig. 2(B)]. In fact, the peak at 1047cm⁻¹ consistently showed up with almost identical intensity for all old TH solution with different concentrations. Figure 2(C) shows the averaged peak intensities at the wavenumber of 764 cm⁻¹ and 1047 cm⁻¹ along with their standard deviations over 10~20 measurements of old TH solution at different locations on the substrates. It clearly shows that as the concentration of TH goes down the peak at 764 cm⁻¹ also goes down (which is the characteristic peak for TH) while the intensity of peak at 1047 cm-1 remains almost the same. Also because of the fact that the peak at 1047cm⁻¹ is absent in all TH solution prepared with fresh DI water , we can conclude that the old water must be contaminated by some chemicals giving a Raman peak at 1047cm$^{-1}$.

As we already know old sample are contaminated by something which can interfere with the SERS experiment, three questions remain to be answered here. What are those contaminants? Where do the contaminants come from? What is the concentration of the contaminants?

3.2 Identification of contaminants with Mass spectrometry

**To answer the first question, we performed mass spectroscopy to indentify the contaminants. The GC-MS spectra showed nine different peaks for the old water sample (Fig. 3). To interpret the mass spectra, the in-built Wiley and NIST libraries were used. The chemicals found are mostly low molecular weight (m/z < 500 Da). The various peaks predicted to correspond to Benzaldehyde, 4-methyl (peak-1 or PK-1) with a mass of 91 Da , Decane, 1-chloro (PK-2) with mass between 43 to 91 Da, 2-Methyltetradecan (PK-3) with mass between 41 to 211 Da, Phenol, 2,4-bis[1,1-dimethylethy-] (PK-8) with a mass 191 Da and Benzoic acid, 4-methyl- (PK-9) with mass between 65 to 136 Da. Since, GC-MS is a fragmenting technique, the subsequent analysis of major peaks revealed that water leachates are heterogenous mixtures of small molecule and Phthalic acid may be the major chemical that is leaching in to the water. It is also well known that Phthalic acid is generally used as a plasticizer to make the plastic flexible[15]**

**{{553 Manas Chanda, Salil K. Roy 2006}}. In addition, Phthalic acid(PA) along with its ester derivatives are widely used as additives in polymer synthesis[16, 17]**

{{544 匿名 2004; 543 Junk,Gregor A. 1974}}. The other chemicals found by GC-MS, may have possibly come from fragmentation of different biocide, slip agent and/or oxidants used in the manufacturing of PP tube (owing to the low molecular weights of those chemicals). We decided to continue our experiment with Phthalic acid as a reference for our subsequent analysis.

3.3 Identification and concentration estimation of contaminants using SERS

**In order to confirm the presence of Phthalic acid (PA) in the leachants we first performed the SERS of PA with different concentrations (Fig. 4 (A)). The key vibrational signs are situated in between 700 and 1700 cm-1 which agrees well with the literatures**[18-21]

{{554 Arenas,J.F. 1980; 555 Colombo,L. 1984; 556 Joo,Sang Woo 2000; 558 Osterrothová,Kateřina 2010}}. The Raman vibrational assignments of PA has been made independently by different authors [above references]. Briefly, the strong peak at 1047 cm-1 is due to C-H wagging, 1141 cm-1 is attributed to C-H bending mode (9a in Wilson notation), 1308 cm-1 is the benzene ring torque mode (3 in Wilson notation) and 1450 cm-1 is the ring breathing vibrational mode (19a in Wilson notation). Further we compare the SERS spectra of fresh Millipore water and old Millipore water stored in PP tube. Fig. 4(B) shows the comparison between the SERS spectra of fresh water, old water and PA solution with the concentration of 1uM and 10uM. Clearly two distinct peaks at 1047 cm-1 and 1141 cm-1 as well as two slightly weak peaks at 1308 cm-1 and 1442 cm-1 showed up in the SERS spectra for old water. While the SERS spectra of fresh water and that of reference background (substrate itself) did not reveal any distinct peaks (Fig. 4 (B)). Due to the similarity in SERS spectra of old water and that of PA characteristic peaks, we believe that the contaminants in old water are having structure similar to PA. The SERS measurement also agrees well with the GC-MS prediction of Phthalic acid as suspected contaminants.  To estimate the concentration of the contaminants, we compare the observed characteristics peak intensity of old water with that of PA with different concentrations. To be more reliable and statistically accurate, we take the average of the intensities from 15~20 measurements at different locations on the SERS substrate for the four characteristic peaks as shown in Fig. 4(C). As shown in Fig. 4 (B) and (C), the Raman peak intensities of old water falls in between those of 1uM and 10uM PA solution. Hence, we can conclude that the derivatives of Phthalic acid in the old water are within the concentration range between 1uM and 10uM.

3.2 Identification and concentration estimation with UV absorption spectroscopy

In addition to GC-MS and SERS measurement, ultraviolet (UV) absorption spectroscopy is also used to confirm and evaluate the suspect contaminants. Compared with SERS and GC-MS measurement, UV absorption works better for quantitative analysis though with lower sensitivity. In our experiment, a quartz cuvette containg the liquid sample is put in Evolution 60 UV-Visible Spectrophotometer for absorption measurement. For calibration of instrumental error, another empty cuvette is used for reference. The scanning wavelength range is from 190nm to 350nm, the scanning step is 0.5nm and the integration time for each step is 1.5 seconds. We measured for fresh water, old water and Phthalic acid solution with different concentrations, shown in Fig. 5 and Fig. S5. The measured UV absorption spectra of Phthalic acid solution correspond to that in literature.22 {{551 Fassinger,W.P. 1959} Comparing old water and fresh water, the absorption of old water is generally higher than fresh water over the whole wavelength range and it shows several peaks. Fresh water gives nearly zero absorbance over the wavelength range. The strong absorbance for old water in the range between 190 nm – 350 nm points to the presence of contaminants which can absorb UV light. We compared the UV absorption spectra of old water and that of Phthalic acid with different concentrations (Fig. 5 (A)). The absorption of Phthalic acid solution increases with the concentration while the shape of the absorption spectra remains the same. The absorption spectrum for Phthalic acid showed three distinct peaks at the wavelength of 200 nm, 236 nm and 282 nm. We found that the absorption spectra of old water is very similar to that of phthalic acid with distinct absorption peaks appearing at 200nm and 236nm respectively. This shows the contaminants in old water, which absorb UV light, have similar structure with Phthalic acid. Above all, the concentration of suspect contaminants in old water falls between 1uM and 10uM because of the fact that the absorption curve for old water situates in between the curves for 1uM PA solution and 10uM PA solution and from the interpolation of the peak intensity for old water we estimate the concentration of suspect contaminants as 2.1 uM. (Fig. 5 (B)).

In conclusion, the UV absorption spectroscopy confirms the chemical identified by GC-MS and SERS measurements that it is the derivatives of Phthalic acid which contaminates the sample and agrees with SERS measurements on the quantitative estimation of the concentration of suspect contaminants to be between 1uM and 10uM.

4. Discussions and future outlook

Even though we have extensively studied the identity of the contaminants leached from PP tubes and managed to estimate their concentration, there are still some questions need to be addressed for more thorough understanding. Also we would like to share and discuss some interesting observations. First of all, the mechanism behind leaching of chemicals from the tube and the dynamics such as how fast do they leach is still unknown. The sample we used for testing is about two months ago but what happened during the first few weeks, days or even hours? As the contaminants we identified are not a single compound, which one will leach first or faster? How do the temperature, pH value and polarity of solvent affects the leaching process? Does the leaching ever reach a saturation and thus the concentration of contaminants no longer increases? All those questions can be systematically investigated with SERS and UV absorption spectroscopy. But they are not covered in this article due to limited time and space.

Secondly, we observed some peaks of Phathlic acid are shifted for different concentrations. Shown in Fig. S4, the peak seen at the wavenumber of $1075cm^{-1}$ for the concentration of 1mM and 100uM is shifted to $1047cm^{-1}$ for the lower concentration of 10uM, 1uM and 100nM. The old water also shows the Raman peak at $1047cm^{-1}$ as Phthalic acid with lower concentrations, shown in Fig. 3(A). Although there is some evidence to show the SERS peaks of Phthalic are shifted in different pH and concentrations [22]

{{548 Tourwé,Els 2006}} the reason for the frequency shift for different concentration here we are inclined to believe is due to difference in types of adsorption. Since the Phthalic acid molecule is adsorbed to the surface of silver substrate after the droplet is dry, pH value does not make sense in this case. But the concentration does make difference for adsorption. If concentration is high, multi-layers of molecules will cover the surface so most molecules are physical adsorbed to the surface (physisorption or weak vander waals force). While with low concentration, the molecules are more likely to form a monolayer on surface thus most molecules will directly contact and form chemicals bond with the silver surface, which is chemical adsorption(chemisorption). It is well studied with charge transfer theory that chemisorption may result in peak frequency shift in the SERS spectra. [23]

{{127 Ricardo Aroca 2006}}

Thirdly, even though SERS spectra and UV absorption spectra for old water match well with those of suspect contaminant Phthalic acid, they are not perfectly identical. In Fig. 2(A) the Raman peaks at $1308cm^{-1}$ and $1422cm^{-1}$ are not so prominent for old water as for Phthalic acid. In Fig. 4(A) we can see the absorption peak around 200nm for old water is at a slightly lower wavelength than that for Phthalic acid. Since we have indentified the contaminants as derivatives of Phthalic acid rather than the Phthalic acid itselves, there may be some differences in the structure of the contaminants and our standard chemicals, both of which are reflected on the SERS and UV absorption spectra.

**Finally, four major peaks at the wavenumber of 1047cm$^{-1}$, 1141cm$^{-1}$, 1308cm$^{-1}$ and 1442cm$^{-1}$ are observed for both old water and Phthalic acid while only the one peak at 1047cm$^{-1}$ is significant for Thiamine Hydrochloride solution in old water, shown in Fig. 2(A). The explanation we come up with for this is that the Raman spectra of mixtures is basically not superposition of Raman spectra of each individual compound, especially when the compounds are polar. For the polar compounds, even though they do not react with each other, they are more likely to interact with each other on molecular level thus changing the conformation, which can be reflected on the Raman spectra. [24]**

{{549 Richards,C.M. 1949}} Neither Phthalic acid nor Thiamine hydrochloride seems to be non-polar here.

5. Conclusion

We found and confirmed that the Polypropylene centrifuge tube may contaminate its liquid contents by leaching chemicals as derivatives of phthalic acid. Even though low in concentration, the contaminants may interfere with the results of some high sensitive analytical measurements such as SERS. GC/MS was used for the identification of the contaminants and with the identity of the suspect contaminants known, SERS and UV absorption spectroscopy were used to confirm the contaminants and to estimate the concentration of them. The measurement results of three spectrometric techniques agree with each other well. With UV absorption spectroscopy, we precisely measured the concentration of derivatives of phthalic acid in water stored in PP tubes for two months as 2.1uM. We propose SERS and UV absorption spectroscopy as two sensitive, simple and quick techniques to detect the contaminants leached from PP tubes.

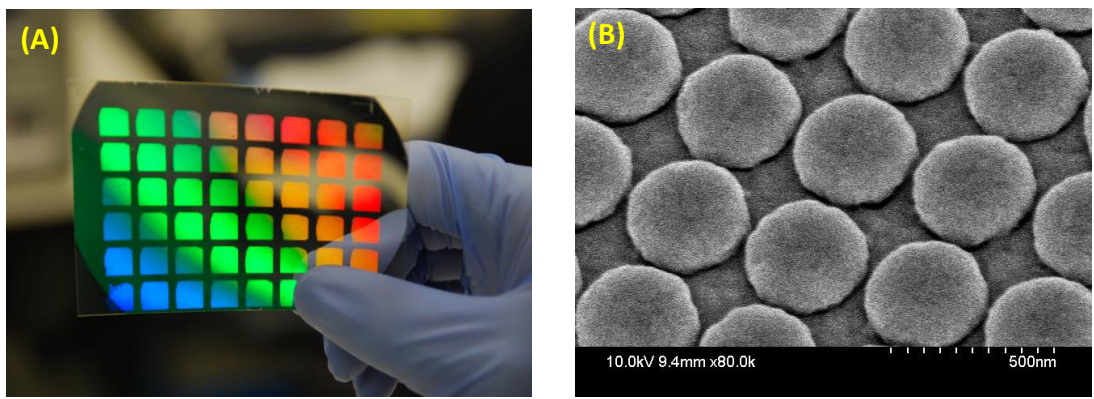

**Figure 1.** (A)Photograph of the nanodome SERS substrate (B) SEM image of the surface of the nanodome SERS substrate with the perspective angle of 25°.

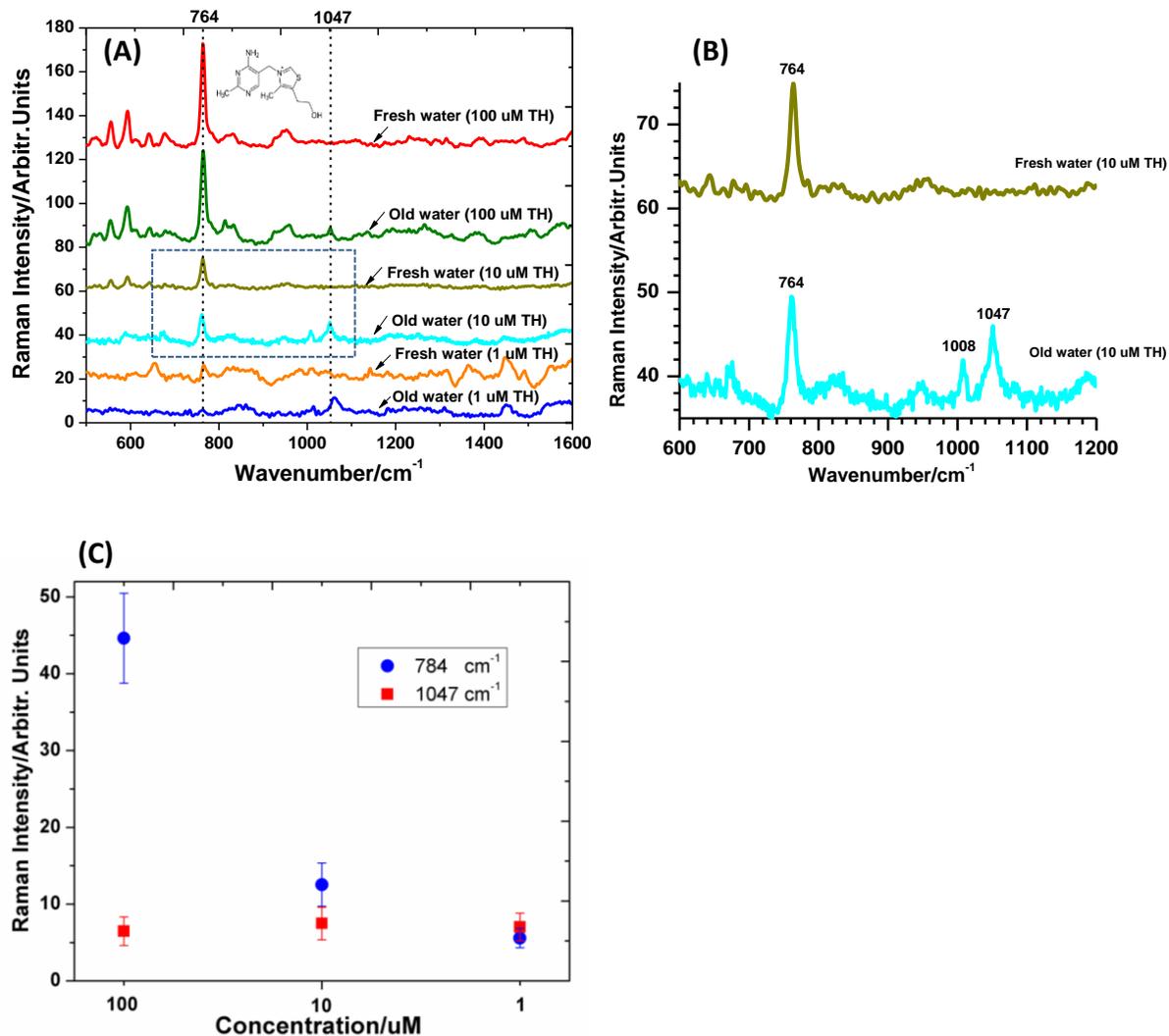

**Figure 2.** (A)SERS spectra of old and fresh Thiamine hydrochloride(TH) solution of the concentrations of 1uM, 100uM and 100uM. (B)Comparison of the SERS spectra for TH solution with the concentration of 10uM (zoomed in image for the cropped region on (A)). (C) Averaged intensities along with the standard deviations of the Raman peaks from old TH solution with 3 different concentrations for the wavenumbers of 764cm⁻¹ and 1047cm⁻¹.

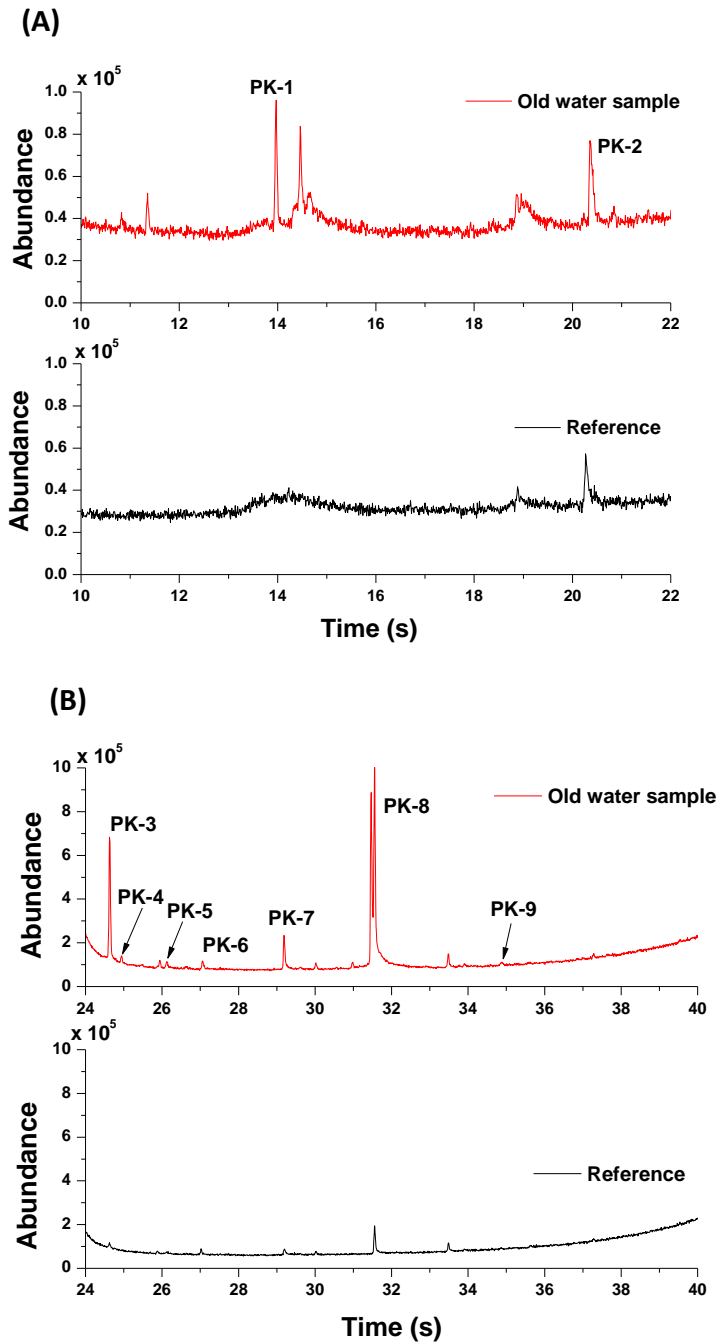

**Fig. 3 (A)** GC-MS spectra of old water showing the two peaks corresponding to contaminants. The details of the contaminants are in the supplementary materials. **(B)** GC-MS spectra of old water showing more peaks at larger retention time corresponding to contaminants.

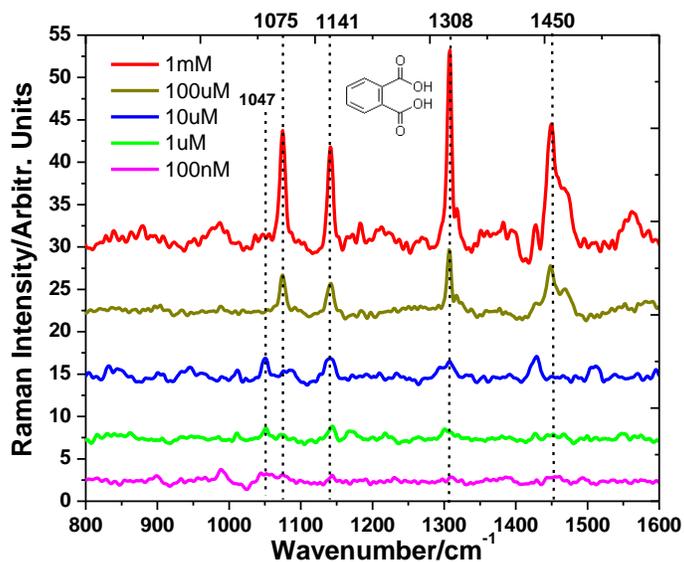

**Fig. 4(A)**

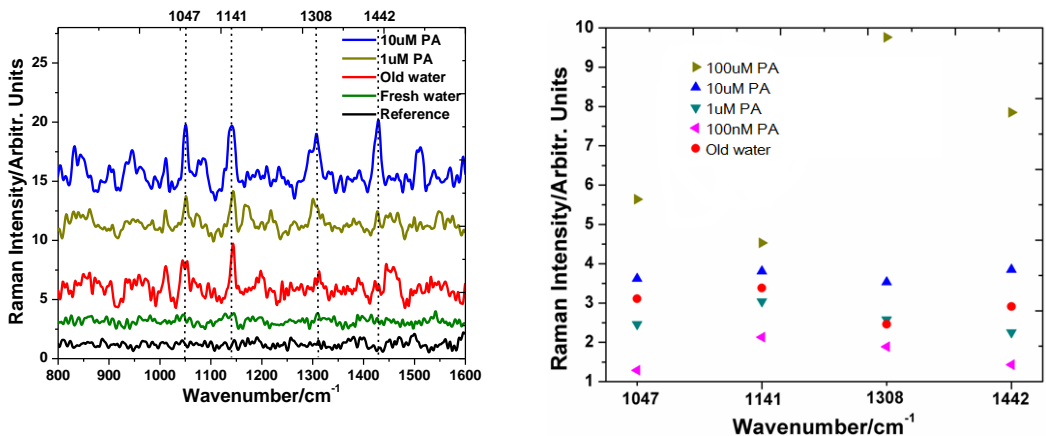

**Figure 4.** (B)SERS spectra of fresh water, old water and Phthalic acid(PA) solution with the concentration of 10uM and 1uM and the bare substrate itself as reference. (C) Averaged peak intensities of old water and Phthalic acid(PA) with different concentrations at the wavenumber of 1047cm-1, 1141cm-1, 1308cm-1 and 1442cm-1.

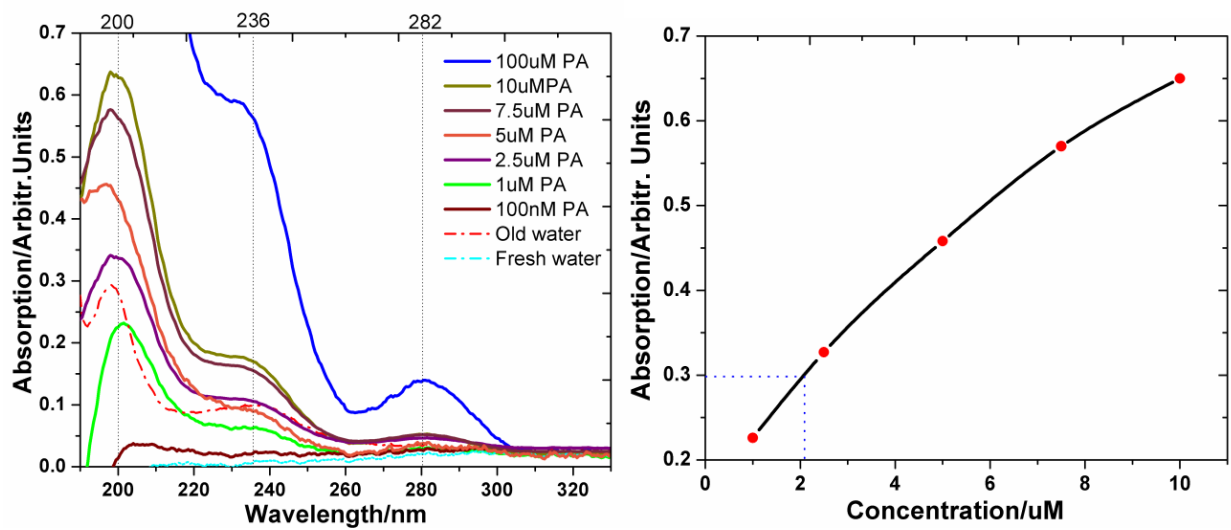

**Figure 5.** (A)Ultraviolet(UV) Absorption spectra for fresh water, old water and Phthalic acid solution with the concentration of 100nM, 1uM, 2.5uM, 5uM, 7.5uM, 10uM, 100uM. (B)Curve fitting to the absorbance at the peak around 200nm on (A) for Phthalic acid with the concentration of 1uM, 2.5uM, 5uM, 7.5uM, 10uM. The dotted line indicates the peak height for old water and its corresponding estimated concentration.